\documentclass[]{elsart}
\usepackage{amssymb}
\usepackage{pst-node,pstricks,graphicx}

\begin{document}                
\begin{frontmatter}
\title{Energy Release in Air Showers}
\author{Markus~Risse\corauthref{cor1}}
\author{and}
\author{Dieter~Heck}
\corauth[cor1]{ {\it Correspondence to}:~M. Risse
(markus.risse@ik.fzk.de)
}

\address{
Forschungszentrum Karlsruhe, Institut f\"ur Kernphysik, 76021 Karlsruhe, Germany 
}


\begin{abstract}                

A simulation study of the energy
released in air due to the development of an extensive air shower
has been carried out using the CORSIKA code.
The contributions to the energy release from different particle
species and energies as well as the typical particle densities 
are investigated.
Special care is taken of particles falling below the energy threshold of the
simulation which contribute about 10\% to the total energy deposition.
The dominant contribution to the total deposition stems from electrons and
positrons from sub-MeV up to a few hundred MeV, with typical transverse
distances between particles exceeding 1 mm for 10 EeV showers.  

\end{abstract}
\end{frontmatter}

\section{Introduction}
\label{sec-intro}

During the shower process initiated by a primary cosmic ray in the atmosphere,
in general, only a small fraction of the initial energy reaches the ground as
high-energy secondary particles. Instead, most of the primary energy is
released in the atmosphere  by ionization and excitation of the air molecules.
A tiny fraction of order $10^{-4}$ is emitted as fluorescence light.
As shown by the Fly's Eye experiment~\cite{flyseye} and its successor,
the High Resolution Fly's Eye~\cite{hires}, the 
fluorescence component of an air shower can be used to detect cosmic rays at energies
exceeding about $10^{18}$~eV. Also the FD detectors of the Pierre Auger 
Observatory~\cite{augerea} 
and planned experiments such as EUSO \cite{euso},
OWL \cite{owl}, and Telescope Array \cite{telarray} are based on this
detection technique.
In contrast to measurements with particle detectors on ground, 
the observation of the longitudinal shower curve in the fluorescence light provides
calorimetric information and therefore conclusions about the primary energy
which are largely independent of the primary particle type and of
unknown details of hadronic interactions at these extreme energies.

In the classical approach of reconstructing the primary energy from
the measured data \cite{song,hires2002},
the amount of light emitted by the shower
at a particular atmospheric depth $X$ (determined from the observables by
``backtracing'' based on detector calibration, atmospheric corrections
and previously performed geometrical reconstruction) is converted to a
number of charged particles $N_{ch}$
assuming a mean ionization rate.
A dependence of the fluorescence yield, i.e. the fraction of the
released energy that is emitted as fluorescence light, on temperature and density
is taken into account according to \cite{kakimoto}.                                        
Finally, $N_{ch} (X)$ is integrated over the path-length and multiplied
by a mean ionization loss rate
and a correction
for missing energy is made.
Air shower simulations are invoked to obtain average values for
the mean ionization rate and for the correction of the missing energy
\cite{song}.

For such primary energy determinations, it is assumed that the amount of
fluorescence photons produced locally per unit length $dN_{fl}/dl$
is proportional to the local energy release $dE_{rel}/dX$ in air,
\begin{equation}
\label{eq-npropedep}
\frac{dN_{fl}}{dl} = y(T,\rho) \cdot \rho_{air}(h) 
			 \cdot \frac{dE_{rel}}{dX} ~~~,
\end{equation}
with $\rho_{air}(h) = dX/dl$ being the atmospheric density, where
$X$ and $l$ are measured along the shower axis.
The fluorescence yield $y(T,\rho)$ of air (in units of emitted photons per
released energy) is a key quantity for evaluating the calorimetric shower
energy.
It has to be determined by laboratory measurements
with individual particles as projectiles \cite{kakimoto,nagano}.
Due to quenching effects, $y$ is dependent on environmental conditions.
Temperature and density have to be varied according to realistic
atmospheric conditions with air as target material,
maybe even taking humidity effects into account.
A considerable experimental effort is started to determine $y(T,\rho)$ more
precisely \cite{badliebenzell}.
Different experiments, some of them at
accelerator facilities, with various measuring conditions are being planned.
Especially, the assumption of a proportionality
between fluorescence light production and deposited ionization energy, which has been
justified to some extent by previous fluorescence yield measurements \cite{kakimoto,nagano}
will be checked thoroughly.

The role of the total deposited ionization energy for the calculation of fluorescence light
makes a precise knowledge of the dominant energy deposit processes in air showers
important.
In Monte Carlo simulations one has to apply a low-energy cutoff for explicit
particle tracking.
Here we introduce a method for calculating the total ionization energy loss
independently of the particular, applied simulation cutoff.

In this work we investigate the characteristics of the
energy release in air showers.
For this purpose, the
CORSIKA simulation code \cite{corsika} has been adapted to quantify
the contributions of different particle species in an air shower
to the energy release. The electromagnetic component as
the dominating one will be examined in more detail.                                       

Our results are also of importance for the planning and interpretation
of laboratory fluorescence yield experiments, as we determine the energy
ranges that dominate the ionization energy deposit in air showers.

The plan of the paper is as follows.
In Chapter~\ref{sec-defedep} it is described how the energy release is
calculated in CORSIKA. This includes the definition of a {\it releasable energy}
for particles which are discarded in the Monte Carlo process 
since their kinetic energy is below the simulation energy threshold.
In Chapter~\ref{sec-longi}, the longitudinal profile of the total
energy release and the contributions of different particle species are
discussed.
The lateral energy deposit distribution together with the densities
of the shower particles are analyzed in Chapter~\ref{sec-lateral}.
The energy spectra of electrons and positrons and their importance for the
total energy release are investigated in Chapter~\ref{sec-spectra}.

\section{Calculation of the energy release with CORSIKA}
\label{sec-defedep}

\subsection{The air shower simulation program CORSIKA}

The  Monte Carlo program package CORSIKA \cite{corsika} is designed
to simulate the development of extensive air showers induced by
various types of primary particles (photons, hadrons, nuclei ... )
in a wide energy range up to the highest energies.
For the particle interactions, external
state-of-the-art codes are employed.
Electromagnetic interactions are simulated using an
adapted version~\cite{upgrade} of the EGS4 code~\cite{egs}, which
includes the Landau-Pomeranchuk-Migdal effect~\cite{lpm}.

To describe hadronic interactions at the energies relevant in this paper,
accelerator data have to be extrapolated by several orders of magnitude
in energy and into a forward kinematic range unobserved in collider experiments.
Therefore, CORSIKA offers a choice of various hadronic interaction models
which differ in predictions of certain air shower characteristics.
For the present investigation, however, the uncertainty related to the
modelling of hadronic interactions turns out to be of minor influence,
as the main characteristics of the energy release in showers
are determined by electromagnetic interactions.
For the calculations presented in the following,
the model QGSJET~01~\cite{qgsjet01,heck_hh} has been
employed for hadronic interactions with energies
E$_{\rm lab} > 80~$GeV, while the GHEISHA routines~\cite{gheisha}
have been used to treat hadronic collisions at lower energies.

To reduce the computational effort in CPU time,
the technique of particle thinning~\cite{hillas},
including weight limitation \cite{kobal,risseicrc}, is applied.
More specifically, a thinning level of $10^{-6}$ has been chosen,
i.e.~no thinning occurs for particles with energies exceeding
$10^{-6}E_0$, with $E_0$ being the primary energy.
In case of thinning, 
the bulk of secondary particles produced in an interaction is
discarded, and only ``representative'' particles are sampled.
A weight factor is assigned to the selected particles to keep energy
conservation~\cite{hillas}.
Upper weight limits of $10^{-6}E_0$/GeV for
electromagnetic particles and $10^{-8}E_0$/GeV for muons and hadrons 
have been chosen to keep artificial fluctuations introduced by
individual particles sufficiently small~\cite{gap-edep} for the
analyses presented in this paper.

In CORSIKA simulations, an (adjustable) energy threshold $E_{thr}$
is adopted for the shower calculation, i.e.~particles are followed
explicitely for $E > E_{thr}$ and discarded for smaller energies.
This is due to the fact that a detailed
calculation down to smallest energies, for instance to the eV-range,
seems both hardly possible (CPU time) and hardly necessary for ``classical''
air shower experiments measuring surviving particles on ground.
In the realm of fluorescence light, however, low-energy particles also contribute
to the ionization and excitation of air molecules.
Therefore, two categories of shower particles are distinguished
for calculating the energy release:
\begin{itemize}
\item particles
      above simulation threshold that 
      are tracked in detail (section~\ref{sec-defioniz}),
\item particles below the simulation threshold
      that are discarded from explicit tracking in the further simulation process
      (section~\ref{sec-defcut}).
\end{itemize}
The final energy release $dE_{rel}/dX$ is the sum of these two contributions,
which will be discussed in the following.

\subsection{Ionization by explicitely tracked charged particles}
\label{sec-defioniz}

The continuous ionization energy loss $dE_i/dx$ of a single charged hadron
or muon traversing matter of thickness $dx$ along its track is calculated
by the Bethe-Bloch stopping power formula 
\begin{equation}
\label{eq-ionizhad}
    \frac{dE_i}{dx}  = \frac{z^2}{\beta^{2}} \kappa_1
             \left(\ln(\gamma^2 -1) - \beta^2 + \kappa_2\right)
\end{equation}
where $\beta = v/c$ is the velocity of the particle in the laboratory
in units of the velocity of light, $\gamma$ is its Lorentz factor, and $z$ is
the charge of the ionizing particle in units of $e$.
The two constants $\kappa_1 = 0.153287~$MeV~g$^{-1}$cm$^{2}$ and
$\kappa_2 = 9.386417 $ are derived from the tables \cite{atomicdata-range}
for dry air.
Additionally the EGS4 routines modified for CORSIKA take into account
the pressure dependent Sternheimer correction~\cite{sternh}.

The transport of electrons and positrons with ionization energy loss 
is treated within CORSIKA in great detail by the EGS4 routines \cite{egs}. 
The contributions of the continuous energy loss stem from the 
soft bremsstrahlung  of photons below the simulation 
threshold and from sub-threshold energy transfer, including ionization
energy loss, to atomic electrons. 
For the latter contribution, EGS4 uses the formulae recommended by 
Berger and Seltzer \cite{berger-seltzer} to apply the concept of 
``restricted stopping power'' \cite{kase-nelson}.

\subsection{Treatment of particles below the simulation threshold}
\label{sec-defcut}
The amount of fluorescence light produced by sub-threshold particles
depends on the reactions the specific particle {\it would} suffer 
on its further way, and therefore on the particle type. 
Antiparticles, for instance, will annihilate and may finally release 
a much larger amount of energy than the (usually quite low) kinetic 
energy.
Therefore, it is useful to define, similar to the concept of
restricted stopping power, the quantity {\it releasable energy} $E_r$
for each particle species. This releasable energy
consists at least of the kinetic energy of the particle.
Its contribution to the total energy release is added during the simulation process 
at the position in air where the particle is discarded.

Stable particles such as electrons 
cannot release more than their kinetic energy ($E_r = E_{kin}$).
Due to annihilation, positrons have a larger releasable energy
($E_r = E_{kin} + 2\cdot m_e = E_{kin} + 1.022$~MeV). However, if the
resulting pair of annihilation photons is above the photon energy threshold,
no energy deposition takes place but
the photons are treated explicitely in the further simulation process.

Muons and mesons are unstable and 
can release also part of their rest mass into ionization.
Some part will be carried away by the decay neutrinos, however.
Antibaryons are assumed to annihilate 
with a nucleon under emission of several pions, so their releasable 
energy, apart from $E_{kin}$, is increased by the anti-baryon rest 
mass and nucleon rest mass. 
In case of unstable particles and the antibaryons, we assumed an effective
fraction of about 1/3 of the releasable energy to be taken into account for
the total energy release, and the remaining 2/3
to be ``lost'' mainly in the neutrino channel.
The simplified treatment of these particle species
below the simulation energy threshold
seems justified as only the contribution of electrons and positrons
below threshold
turns out to be significant for the present analysis.

The assumption of locality of the energy deposited by the sub-threshold particles
is valid for conventional threshold settings in the simulation~\cite{gap-edep}.
While, for instance, a usual choice of the electron energy threshold is about 0.1~MeV,
even for relatively large threshold values in the MeV range
the particles would be stopped within a depth of only a few g~cm$^{-2}$.

\section{Longitudinal development}
\label{sec-longi}

Air showers initiated by different primary particle types
(proton, iron nuclei, and photons)
have been calculated with various primary energies ($10^{18} - 10^{20}$~eV)
and shower zenith angles.
The main conclusions discussed in the following, however, are largely independent
of the primary parameter choice and also of the shower-to-shower fluctuations
(see also~\cite{risse_icrc03,nerling}).
The distributions given in the Figures illustrate the case of 
an iron-induced event of primary energy $10^{19}$~eV with $45^{\circ}$ inclination.

\begin{figure}[t]
\begin{center}
\includegraphics[height=9.5cm,angle=0]{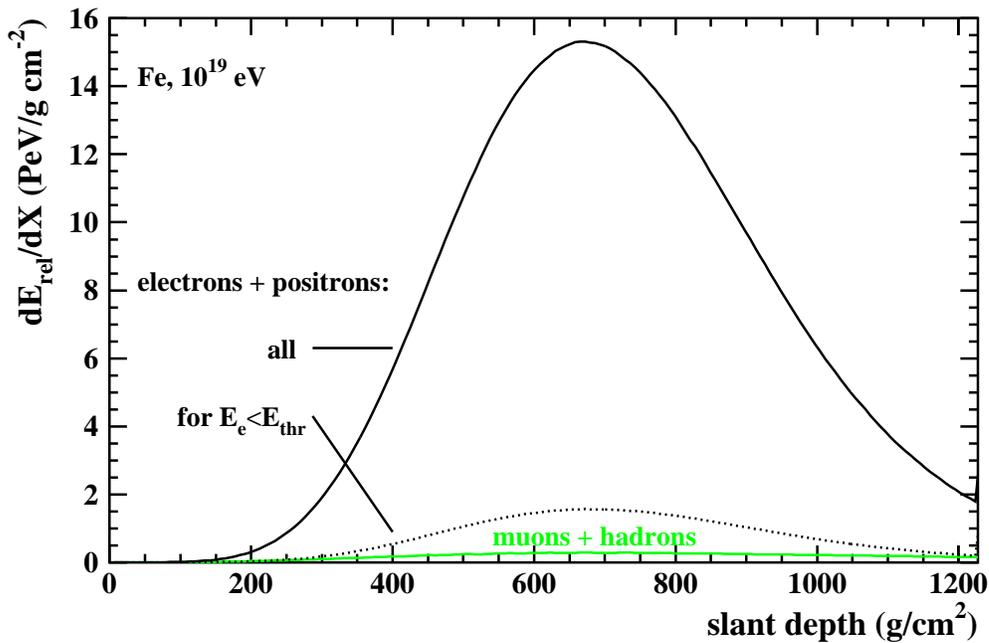}
\caption{Longitudinal shower development:
	 Energy release of electrons plus positrons and of muons plus hadrons.
	 Additionally, the contribution of electromagnetic particles
	 below the simulation energy threshold $E_{thr}=0.1$~MeV is given.
        }
\label{fig-longi.div}
\end{center}
\end{figure}
The longitudinal development of the energy release $dE_{rel}/dX$ of the shower
in air is presented in Figure~\ref{fig-longi.div}.
The definition of the path length $dX$ deserves explanation.
Plotted is the energy release in the layer of air between slant depths $X$ and $X+dX$,
i.e.~in the direction of the shower axis. A particle propagating with a non-zero
angle $\vartheta$ towards the shower axis thus travels through an effective amount
of matter of $dx = dX/\cos(\vartheta)$ while traversing this slant depth interval.
The corresponding larger energy loss, or equivalently the increased
total tracklength of the shower particles due to their angular spread,
is taken into account in CORSIKA.
This effect has also been pointed out recently in \cite{muniz}.

In Figure~\ref{fig-longi.div}, different contributions to the energy release are displayed.
As expected, the main contribution stems from 
electrons and positrons, the most numerous charged particles.
Around shower maximum, less than 2-3\% are resulting from muons and hadrons.
Thus, electromagnetic particles should be the main target for the study of
energy release.

A fraction of about 10\% to the energy release is due to discarded
electrons and positrons below a simulation energy threshold of 0.1~MeV.
This value is in agreement with findings in~\cite{song}.
High-energy electrons produce via M{\o}ller scattering a significant number of
low-energy electrons.
Additionally, the many low-energy shower photons transfer a considerable
energy fraction by Compton scattering to low-energy electrons.

A closer inspection shows that the development of the hadronic and muonic parts
differ somewhat from the electromagnetic ones: At very early (and also late)
development stages the electromagnetic fraction to the local energy release
is decreased. For the current analysis, this is of minor interest, however,
as the fluorescence light production at these shower stages is marginal
compared to the maximum region.

It is important to point out that
the shower profile of the total energy release is not influenced when
varying the simulation energy threshold within reasonable limits, as
also shown in~\cite{gap-edep}.
Adopting, for instance, higher thresholds would
increase the contribution of the discarded particles, but this would just be balanced
by a decreased contribution of the explicitely tracked particles. 

\section{Lateral spread and particle densities}
\label{sec-lateral}

For laboratory measurements of the fluorescence yield,
the question arises whether the particle densities in air
showers are small enough to allow for an undisturbed
de-excitation or whether an ionized or excited air molecule
might be influenced by another nearby shower electron.
For this reason, the lateral spread of the energy release 
and the corresponding density of electrons and positrons as the
component which dominates the energy release, is studied.
More detailed investigations on the lateral distribution of the
energy release in showers including the possibility of observing
the shower width with fluorescence telescopes are given 
elsewhere~\cite{gora}; here we focus on questions relevant
to laboratory measurements of the fluorescence yield.

\begin{figure}[t]
\begin{center}
\includegraphics[height=14cm,angle=0]{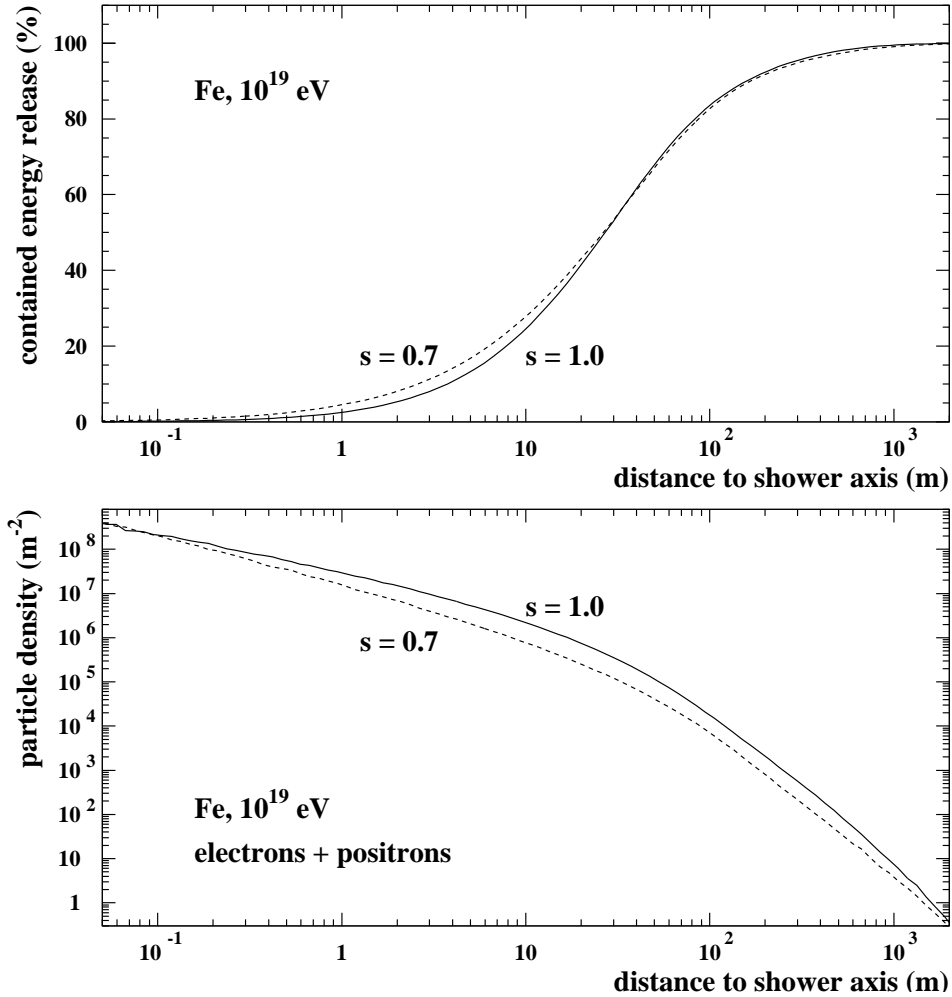}
\caption{Upper panel: Energy release contained within a given distance of the shower axis
versus the distance at shower ages $s = 0.7$ and $s = 1.0$.
Lower panel: Lateral distribution of electrons plus positrons.}
\label{fig-latdens}
\end{center}
\end{figure}
In Figure~\ref{fig-latdens} the amount of released energy contained
within a given distance to the shower axis is plotted as a function
of the distance (upper panel) for different values of the 
shower age $s$, defined as $s = 3X/(X+2X_{max})$. Also given is the density in particles
per square-meter of electrons and positrons
(lower panel). As the lifetime of the radiating molecular states is of the
order 30$-$70~ns and thus comparable to the traversal time of the shower for
a given air particle (or correspondingly to the shower ``thickness''), 
transverse particle distances are regarded for a conservative estimate of the particle
separation.

Particles with core distances below 1~m contribute only little to the energy release:
Though the particle densities are largest here, due to phase space the absolute
particle number is comparatively small.
The main energy release of about 80$-$85\% occurs at distances between $1 -100$~m
from the shower axis.
For core distances around 20~m, a typical transverse particle separation of
$\simeq 1$~mm for 10~EeV 
showers is obtained. This value holds for the shower maximum, at other
stages of the shower development observable in the flourescence light
the particle separations are larger.
The densities are roughly scaling with the primary energy, while the dependence on
the primary particle type is of minor importance in this context.
For 100~EeV showers at relatively small core distances of about 3~m, for
instance, the average transverse particle separation amounts to $\simeq 0.1$~mm.
Thus, with respect to the size of the ionization region around the charged
particles in air (usually $<$1~$\mu$m), 
this is a large separation resulting in a relatively ``undisturbed''
de-excitation of the air molecules. 
High-density particle bunches should therefore be avoided
in fluorescence yield measurements as the fluorescence yield,
apart from extreme cases of high shower particle densities,
would be obtained in conditions not typical for fluorescence
light emission in air showers.

\section{Energy spectra}
\label{sec-spectra}

\begin{figure}[t]
\begin{center}
\includegraphics[height=14cm,angle=0]{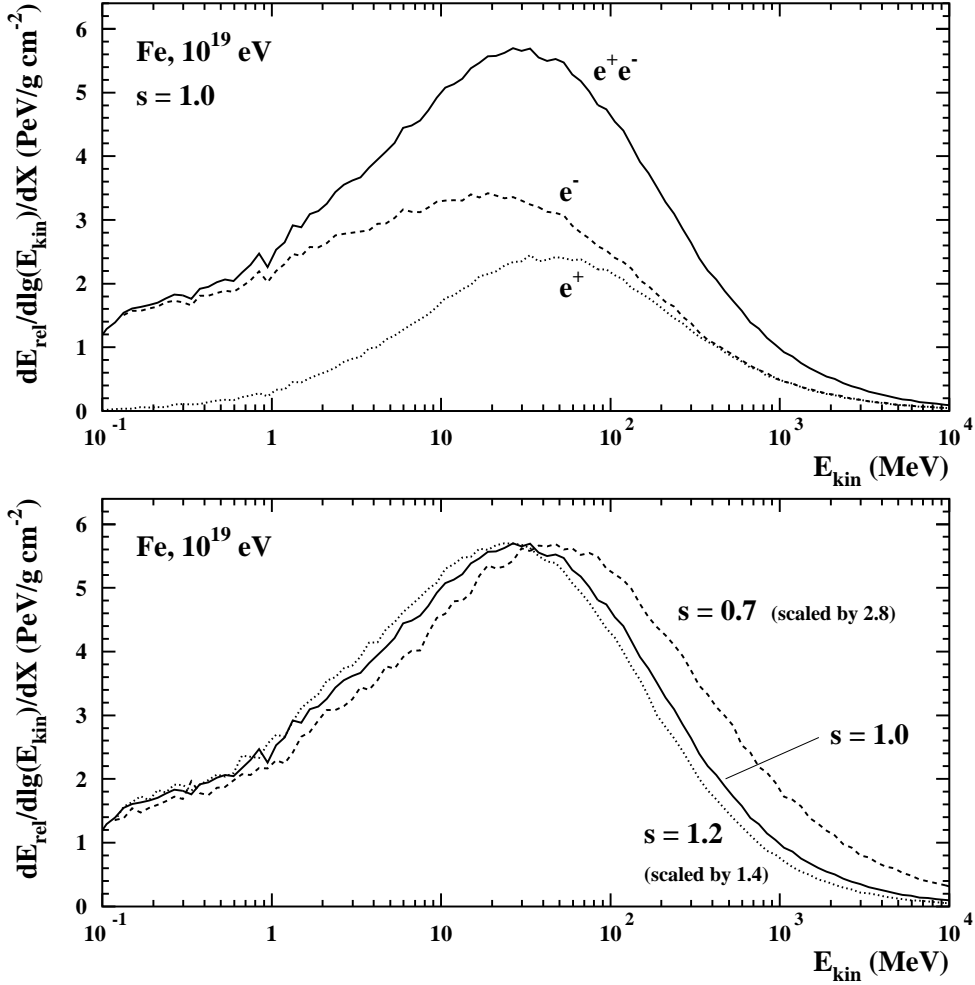}
\caption{Contribution to the energy release per matter traversed in shower direction
as a function of the kinetic particle energy. Simulation for primary iron,
$10^{19}$~eV. Upper panel: Individual and combined contributions of electrons and
positrons at shower maximum.
Lower panel: Combined contribution of electrons and positrons for different
shower ages (normalized to the same height of maximum).}
\label{fig-edep.espek}
\end{center}
\end{figure}

The particle energies contributing to the energy release, an important
input quantity for the layout of fluorescence yield measurements,
are shown in Figure~\ref{fig-edep.espek}. For this graph, only electrons
and positrons are taken into account, since they dominate
the energy release.
The main contribution comes from particles 
with energies below 1~GeV, with a tail towards small energies.
A broad maximum is visible at particle energies around 10$-$50~MeV,
below the critical energy of electrons in air ($\simeq$84~MeV).

The spectral shape mainly reflects the
particle energy spectrum~\cite{risseicrc}. Especially the contributions of the
lower energies are more pronounced, however.
This is due firstly to the increased specific energy loss (Bethe-Bloch formula),
and secondly to a larger average path length $dx = dX/\cos(\vartheta)$ through the
slant depth interval $dX$, since at lower electron energies the dispersion of particle
angles is increasing.
As a guideline, in Table~\ref{tab-edepenerg} the contribution
to the electromagnetic energy release around shower maximum
is estimated for different ranges in kinetic energy.
The value for \mbox{$E_{kin} < 0.1$~MeV} is given by the
contribution of the particles discarded due to the 
simulation energy threshold.

\begin{table}[b]
\begin{center}
\caption{Estimates for the contribution of different ranges in
kinetic energy to the energy release by electrons and positrons.
The uncertainty of the values is about $\pm$2 (in~\%).}
\vskip 0.5 cm
\label{tab-edepenerg}
\begin{tabular}{|c||c|c|c|c|c|c|}
\hline
Energy in MeV      & $<0.1$ & 0.1-1 & 1-10 & 10-100 & 100-1000 & $>$1000 \\ \hline
Contribution in \% & 10     &   12  &  23  &   35   &    17    &    3    \\ \hline
\end{tabular}
\end{center}
\end{table}

While at higher kinetic energies
($E_{kin} > 300$~MeV) electrons and positrons contribute about
equally to the energy release, at lower energies only electrons survive
due to positron annihilation. The annihilation photons
will eventually transfer the energy to electrons by Compton scattering.

The range of particle energies that mainly contribute to the energy release,
is to a good approximation
quite independent of the primary particle type (including primary photons)
and primary energy. 
Also the dependence on shower age is small, as shown in the lower panel
of Figure~\ref{fig-edep.espek}.
For instance, at earlier development stages before the shower maximum,
the contribution is only slightly shifted to higher electron energies.
These results may be understood, since the particle energy spectrum
is known to show a small, but in this context only minor dependence
on the primary particle type and on shower age~\cite{risseicrc,nerling}.

\section{Conclusion}
The presented method of treating sub-threshold particles allows the precise
calculation of the total energy deposit.
For shower calculations, the energy release provided by CORSIKA
can be transformed to fluorescence light based on existing and upcoming
fluorescence yield measurements.

The energy release in air showers has also been studied with respect to
currently planned laboratory fluorescence yield measurements.
Most relevant is the
determination of the yield for electrons and positrons with energies
in the range from sub-MeV up to a few hundred MeV. The typical lateral
particle separation
is relatively large with 1~mm or more for 10~EeV showers at shower
distances where most energy is released and thus, presumably,
the main fluorescence light production occurs.

Finally, we want to mention that based on the concept of total shower size
one can define the mean energy deposit per particle~\cite{flyseye,song}.
This quantity has been subject of many discussions and will be studied
in detail in a forthcoming publication.

{\it Acknowledgements.} The useful discussions with 
D.~G\'ora, H.~Klages, J.~Knapp, P.~Sommers, and T.~Wal\-denmaier
and in particular with R.~Engel
are gratefully acknowledged.



\begin{thebibliography}{99}

\bibitem{flyseye} R.M.~Baltrusaitis et al., {\it Nucl. Instr. Meth.} {\bf A240} (1985) 410

\bibitem{hires} T.~Abu-Zayyad et al., {\it Nucl. Instr. Meth.} {\bf A450} (2000) 253

\bibitem{augerea} J.~Abraham et al., Auger Collaboration, submitted to 
                   {\it Nucl. Inst. Meth.} (2003)

\bibitem{euso} L.~Scarsi et al., {\it $27^{th}$ Int.~Cosmic Ray Conf.}, Hamburg
{\bf 2} (2001) 839

\bibitem{owl} L.~Scarsi et al., {\it $26^{th}$ Int.~Cosmic Ray Conf.}, Salt Lake City
{\bf 2} (1999) 384

\bibitem{telarray} T.~Aoki et al., {\it $27^{th}$ Int.~Cosmic Ray Conf.}, Hamburg
{\bf 2} (2001) 915

\bibitem{song}   C.~Song, Z.~Cao, B.R.~Dawson, B.E.~Fick, P.~Sokolsky,
		 and X.~Zhang, {\it Astropart. Phys.} {\bf 14} (2000) 7

\bibitem{hires2002} T. Abu-Zayyad et al., HiRes Collaboration, submitted to
		   {\it Astropart. Phys.} (2003); preprint astro-ph/0208301 (2002)

\bibitem{kakimoto} F.~Kakimoto, E.C.~Loh, M.~Nagano, H.~Okuno, M.~Teshima,
		 and S.Ueno, {\it Nucl. Instr. Meth.} {\bf A 372} (1996) 527

\bibitem{nagano} M.~Nagano, K.~Kobayakawa, N.~Sakaki, and K.~Ando,
                 {\it Astropart.~Phys.} (2003) in press; preprint astro-ph/0303193

\bibitem{badliebenzell} International workshops in Utah (USA) 2002 and Bad Liebenzell 
		 (Germany) 2003;
		 see www.physics.utah.edu/\~{}fiwaf/done ~and~ www.auger.de/events

\bibitem{corsika}D.~Heck, J.~Knapp, J.N.~Capdevielle, G.~Schatz, and
		 T.~Thouw, Report {\bf FZKA 6019}, Forschungszentrum
		 Karls\-ruhe (1998); www-ik.fzk.de/\~{}heck/corsika

\bibitem{upgrade} D.~Heck and J.~Knapp, Report {\bf FZKA 6097}, Forschungszentrum
		 Karls\-ruhe (1998)

\bibitem{egs}    W.R.~Nelson, H.~Hirayama, and D.W.O.~Rogers, Report
		 {\bf SLAC 265}, Stanford Linear Accelerator Center (1985)

\bibitem{lpm} L.D.~Landau and I.Ya.~Pomeranchuk, {\it Dokl. Akad. Nauk SSSR} {\bf 92}
	       (1953) 535 \& 735 (in Russian);
	       A.B.~Migdal, {\it Phys. Rev.} {\bf 103} (1956) 1811

\bibitem{qgsjet01} N.N.~Kalmykov, S.S.~Ostapchenko, and A.I.~Pavlov,
	      {\it Nucl.~Phys.~B (Proc.~Suppl.)} {\bf 52B} (1997) 17

\bibitem{heck_hh} D.~Heck et al., {\it Proc. $27^{th}$ Int. Cosmic Ray Conf.},
                   Hamburg (Germany) {\bf 1} (2001) 233

\bibitem{gheisha} H.~Fesefeldt, Report {\bf PITHA-85/02}, RWTH Aachen (1985)

\bibitem{hillas} A.M.~Hillas, {\it Nucl.~Phys.~B (Proc.~Suppl.)} {\bf 52B} (1997) 29

\bibitem{kobal} M.~Kobal, Pierre Auger Collaboration, {\it Astropart.~Phys.}
		{\bf 15} (2001) 259

\bibitem{risseicrc} M.~Risse, D.~Heck, J.~Knapp, and S.S.~Ostapchenko, 
                {\it Proc.~$27^{th}$ Int.~Cosmic Ray Conf.}, Hamburg (Germany), 
                {\bf 2} (2001) 522

\bibitem{gap-edep} M.~Risse and D.~Heck, {\it Auger Internal Note} {\bf GAP-2002-043},
                www.auger.org, (2002)

\bibitem{atomicdata-range} R.M.~Sternheimer, M.J.~Berger and S.M.~Seltzer,
                 {\it Atomic Nucl. Data Tables} {\bf 30} (1984) 261

\bibitem{sternh} R.M.~Sternheimer et al., {\it Phys. Rev.} {\bf B26}
		 (1982) 6067

\bibitem{berger-seltzer}M.J.~Berger and S.M.~Seltzer, Report {\bf
                 NASA-SP-3012} (1964)

\bibitem{kase-nelson} K.R.~Kase and W.R.~Nelson, {\it Concepts of
                  Radiation Dosimetry}, Pergamon Press, New York (1979)

\bibitem{risse_icrc03} M.~Risse and D.~Heck, {\it Proc.~$28^{th}$ Int.~Cosmic
		  Ray Conf.}, Tsukuba (Japan) (2003) 607

\bibitem{nerling} F.~Nerling, R.~Engel, C.~Guerard, L.~Perrone, and M.~Risse,
             {\it Proc.~$28^{th}$ Int.~Cosmic Ray Conf.}, Tsukuba (Japan) (2003) 611

\bibitem{muniz} J.~Alvarez-Mu\~niz, E.~Marqu\'es, R.A.~V\'azquez, and E.~Zas,
	       {\it Phys.~Rev.~D} {\bf 67} (2003) 101303

\bibitem{gora} D.~G\'ora, P.~Homola, M.~Kutschera, J.~Niemiec, B.~Wilczy\'nska, and
               H.~Wil\-czy\'nski, {\it Astropart.~Phys.} {\bf 16} (2001) 129;
               D.~G\'ora et al., in preparation (2003)

\end{thebibliography}
\end{document}